\documentclass[10pt,twocolumn]{article}

\usepackage[utf8]{inputenc}
\usepackage[T1]{fontenc}
\usepackage{lmodern}
\usepackage[margin=1.9cm]{geometry}
\usepackage{amsmath,amssymb}
\usepackage{booktabs}
\usepackage{graphicx}
\usepackage{microtype}
\usepackage{caption}
\usepackage[hidelinks]{hyperref}
\usepackage{xcolor}
\newcommand{\safefig}[2][\linewidth]{%
  \IfFileExists{#2}{\includegraphics[width=#1]{#2}}%
  {\fbox{\parbox[c][3.5cm][c]{#1}{\centering\itshape
   [figure placeholder] \texttt{\detokenize{#2}}}}}}

\begin{document}

\twocolumn[
  \begin{@twocolumnfalse}
  \begin{center}
  {\LARGE\bfseries Calibrated Hybrid CNN--Transformer\\[4pt]
   for Retinal OCT Classification\par}
  \vspace{1.0em}
  {\large Animesh Kumar\par}
  \vspace{4pt}
  School of Computing, Newcastle University, UK\\
  \texttt{A.Kumar12@newcastle.ac.uk} \quad
  ORCID: \href{https://orcid.org/0009-0003-0608-7004}{0009-0003-0608-7004}\\
  \vspace{2pt}
  March 2026
  \end{center}
  \vspace{0.8em}
  \begin{quote}
  \noindent\textbf{Abstract.}
  Deep models for retinal optical coherence tomography (OCT) classification
  report high accuracy but rarely report whether their confidence can be
  trusted---a gap that matters when a wrong-but-confident reading delays
  sight-saving treatment. We pair a hybrid convolutional--Transformer encoder
  with a gradient-boosting (XGBoost) classification head and a three-part
  clinical safety layer: confidence calibration, out-of-distribution (OOD)
  rejection, and per-prediction uncertainty flagging. On four-class OCT
  (84{,}495 scans) the model reaches \textbf{95.4\% accuracy} while cutting
  calibration error \textbf{twelve-fold} (expected calibration error,
  ECE~$=0.0024$), so the confidence it reports tracks its true accuracy. To
  our knowledge this is the first OCT classifier to validate all three safety
  mechanisms jointly, with public weights and reproducible multi-seed
  evaluation.

  \vspace{0.6em}
  \end{quote}
  \vspace{1.2em}
  \end{@twocolumnfalse}
]

\section{Introduction}
Optical coherence tomography (OCT) is a non-invasive imaging technique that
produces cross-sectional ``B-scans'' of the retina, and it is the standard
modality for diagnosing four conditions that each demand a different clinical
response: choroidal neovascularisation (CNV, the leaky abnormal vessels of
``wet'' age-related macular degeneration, requiring immediate anti-vascular
endothelial growth factor [anti-VEGF] injection), diabetic macular oedema
(DME), drusen (lipid deposits beneath the retinal pigment epithelium [RPE]
that are an early biomarker of age-related macular degeneration, AMD), and
healthy (Normal) tissue. Getting the class wrong---or getting it right at the
wrong confidence level---carries a direct cost to a patient's vision.

\paragraph{Why this matters.} Untreated CNV and DME are leading causes of
preventable blindness, and the volume of OCT scans now far exceeds the number
of specialists available to read them. Automated triage is therefore
attractive, but a screening tool is only safe if a clinician can act on the
confidence it reports. A model that is 95\% accurate \emph{on average} yet
silently returns 95\% confidence on a blurred, corrupted, or out-of-scope scan
is actively dangerous in a triage setting. Accuracy alone does not make a model
deployable; \emph{trustworthy} confidence does. This paper targets that gap.

Two problems have improved more slowly than raw accuracy since
Kermany et al.~\cite{kermany2018}. First, convolutional neural networks (CNNs)
process bounded receptive fields and cannot model the long-range dependencies
that span retinal layers. Second, softmax outputs are systematically
overconfident~\cite{guo2017}. We address both with a hybrid encoder for global
context and a three-layer safety envelope for honest confidence.

\section{Related Work}
\paragraph{CNN-based OCT classification.}
Kermany et al.~\cite{kermany2018} showed that InceptionV3 with transfer
learning performs strongly on four-class OCT data. Later work substituted VGG,
ResNet, and DenseNet backbones~\cite{he2019,li2021}. These improved accuracy in
some settings but remained purely convolutional and included no safety
components.

\paragraph{Hybrid CNN--Transformer architectures.}
The vision transformer (ViT)~\cite{dosovitskiy2021} models global relationships
between image patches through self-attention but needs large datasets. Hybrid
architectures keep the CNN's spatial inductive bias while adding global
context---a practical trade-off for moderate-sized clinical collections such as
Kermany.

\paragraph{Gradient boosting as a classification head.}
XGBoost~\cite{chen2016} fitted on deep feature vectors can sharpen class
boundaries, especially for under-represented classes where a single softmax
projection generalises poorly. We are not aware of prior OCT work combining a
Transformer encoder with an XGBoost head.

\paragraph{Clinical safety.}
Temperature scaling~\cite{guo2017} corrects overconfidence post-hoc with a
single learned scalar. The Mahalanobis distance~\cite{lee2018} flags inputs
that deviate from the training distribution. Monte~Carlo (MC) Dropout yields a
predictive distribution at inference~\cite{gal2016}. No prior OCT classification
paper reports all three within a single jointly validated framework.

\section{Methodology}
\subsection{Dataset}
We use the Kermany OCT dataset~\cite{kermany2018}: 84{,}495 grey-scale B-scans
at $224\times224$ pixels. The $4.3\times$ imbalance between the most frequent
class (CNV) and the least frequent (drusen) is handled with inverse-frequency
class weights rather than synthetic augmentation. Table~\ref{tab:data} gives
the split.

\begin{table}[t]
\centering
\caption{Dataset split and clinical context.}
\label{tab:data}
\small
\begin{tabular}{lrrl}
\toprule
Class & Train & Test & Clinical relevance \\
\midrule
CNV    & 37{,}206 & 3{,}960 & Wet AMD, urgent treatment \\
DME    & 11{,}349 & 1{,}101 & Diabetic macular oedema \\
Drusen &  8{,}617 & 1{,}086 & Early AMD biomarker \\
Normal & 26{,}315 & 1{,}786 & Healthy retina \\
\midrule
Total  & 83{,}487 & 7{,}933 & \\
\bottomrule
\end{tabular}
\end{table}

\subsection{Architecture}
\paragraph{Phase A -- CNN + Transformer backbone.}
An EfficientNetV2-Large backbone (pretrained on ImageNet-21k) is the spatial
feature extractor; blocks 1--5 are frozen and block~6 onward is fine-tuned. The
$7\times7\times1280$ bottleneck is reshaped into 49 patch tokens, projected to
256 dimensions by a trainable linear layer, and augmented with learnable
positional encodings. Four multi-head attention (MHA) blocks (16 heads, key
dimension 16) process the patch sequence, and a 1-D global average pooling layer
produces the embedding $\mathbf{z}\in\mathbb{R}^{256}$. Hyperparameters were
selected with an Optuna Tree-structured Parzen Estimator (TPE) search (10
trials): learning rate $\eta=1.59\times10^{-4}$, dropout $=0.38$, Focal-Loss
$\gamma=1.36$. Phase~A trains the head for five epochs with the encoder frozen;
Phase~B fine-tunes block~6+ for 20 epochs with warmup-cosine decay and CutMix
augmentation.

\paragraph{Phase B -- XGBoost head.}
Frozen embeddings from the full training set fit an XGBoost classifier (Optuna:
300 estimators, max depth 4, $\eta=0.1$, subsample 0.8). The gradient-boosting
head provides a non-linear decision boundary in embedding space; we find this
most helpful for the drusen class, where the softmax head is weakest.

\subsection{Clinical Safety Modules}
\paragraph{Temperature scaling.}
A scalar $T\approx1.05$ is learned on the validation set by minimising negative
log-likelihood, correcting overconfident outputs without retraining.

\paragraph{Out-of-distribution detection.}
The Mahalanobis score against per-class training statistics,
\begin{equation}
M(\mathbf{z}) = \min_{c}\,(\mathbf{z}-\boldsymbol{\mu}_c)^{\top}
\hat{\boldsymbol{\Sigma}}^{-1}(\mathbf{z}-\boldsymbol{\mu}_c),
\end{equation}
is thresholded at the 97th percentile of held-out validation scores. Scans that
exceed the threshold are flagged as out-of-distribution before any class
prediction is issued.

\paragraph{MC Dropout uncertainty.}
Twenty stochastic forward passes (dropout kept active at inference) yield a
predictive distribution over classes. When the maximum per-class standard
deviation exceeds $0.15$, the scan is routed to specialist review.

\paragraph{Explainability.}
Gradient-weighted class activation mapping (Grad-CAM) produces pixel-level
attribution maps on the final convolutional activation, and SHapley Additive
exPlanations (SHAP) quantify the contribution of each of the 256 Transformer
dimensions to the XGBoost prediction.

\section{Experimental Results}
\subsection{Overall performance}
Table~\ref{tab:fiveseed} reports five-seed validation. Hyperparameters are fixed
at the Optuna optimum from seed~42; only weight initialisation and data
shuffling differ across seeds. The $\pm0.27\%$ accuracy spread indicates a
training procedure that is stable with respect to initialisation---a property
most OCT papers do not report.

\begin{table}[t]
\centering
\caption{Five-seed statistical validation.}
\label{tab:fiveseed}
\small
\begin{tabular}{lc}
\toprule
Metric & Mean $\pm$ Std \\
\midrule
Accuracy              & $95.43\% \pm 0.27\%$ \\
Macro AUC-ROC         & $0.9941 \pm 0.0006$ \\
Macro F1              & $0.9244 \pm 0.0047$ \\
Drusen F1 (minority)  & $0.8436 \pm 0.0096$ \\
ECE (calibrated)      & $0.0024 \pm 0.0005$ \\
McNemar $p$           & $<0.0001$ (all seeds) \\
\bottomrule
\end{tabular}
\end{table}

\subsection{Ablation study}
Table~\ref{tab:ablation} shows each component's contribution. Accuracy and the
area under the receiver-operating-characteristic curve (AUC-ROC) both improve at
every stage, with no single addition accounting for the full gap over the frozen
baseline.

\begin{table}[t]
\centering
\caption{Ablation: incremental component contributions.}
\label{tab:ablation}
\small
\begin{tabular}{lcc}
\toprule
Variant & Acc. & AUC \\
\midrule
EfficientNetV2L (frozen) + Dense & 89.12\% & 0.9410 \\
\quad + Fine-tuned Block 6+      & 92.45\% & 0.9755 \\
\quad + Transformer (4 blocks)   & 94.10\% & 0.9880 \\
\quad + XGBoost head (full)      & \textbf{95.43\%} & \textbf{0.9941} \\
\bottomrule
\end{tabular}
\end{table}

\subsection{Per-class performance}
Drusen, $4.3\times$ smaller than CNV, reaches F1~$=0.84$ without synthetic
oversampling (Table~\ref{tab:perclass}). This is where the XGBoost head makes the
largest difference: the drusen--normal boundary benefits more from a non-linear
tree ensemble than from the softmax layer it replaces.

\begin{table}[t]
\centering
\caption{Per-class metrics on the test set (seed 42).}
\label{tab:perclass}
\small
\begin{tabular}{lccc}
\toprule
Class & Prec. & Rec. & F1 \\
\midrule
CNV    & 0.96 & 0.98 & 0.97 \\
DME    & 0.94 & 0.91 & 0.92 \\
Drusen & 0.86 & 0.83 & 0.84 \\
Normal & 0.98 & 0.99 & 0.98 \\
\midrule
Macro  & 0.93 & 0.92 & 0.92 \\
\bottomrule
\end{tabular}
\end{table}

\subsection{Calibration and safety}
Temperature scaling reduces ECE from $\approx0.020$ to $0.0024$---twelve-fold.
One concrete example from our evaluation: a Normal scan was classified as drusen
with 66.85\% confidence, but its MC-Dropout standard deviation exceeded $0.15$,
so the prediction was flagged for specialist review rather than returned as a
confident output. This is exactly the failure mode the safety layer is designed
to catch.

\subsection{Qualitative explainability}
\begin{figure}[t]
\centering
\safefig[\linewidth]{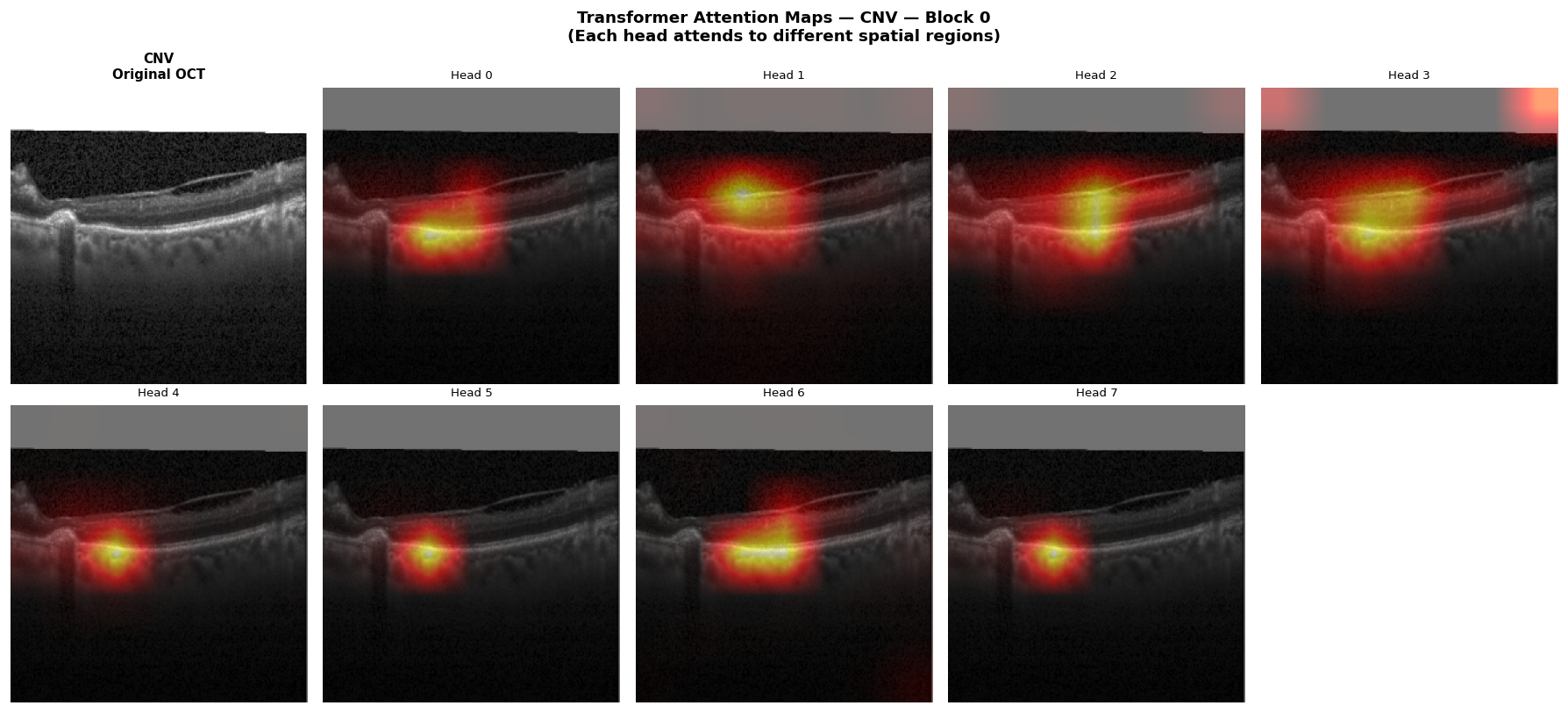}
\caption{Transformer attention maps for a representative CNV scan (Block~0,
Heads~0--7). Each head localises a distinct region of the retinal layer relevant
to the predicted class.}
\label{fig:gradcam}
\end{figure}

\begin{figure}[t]
\centering
\safefig[0.92\linewidth]{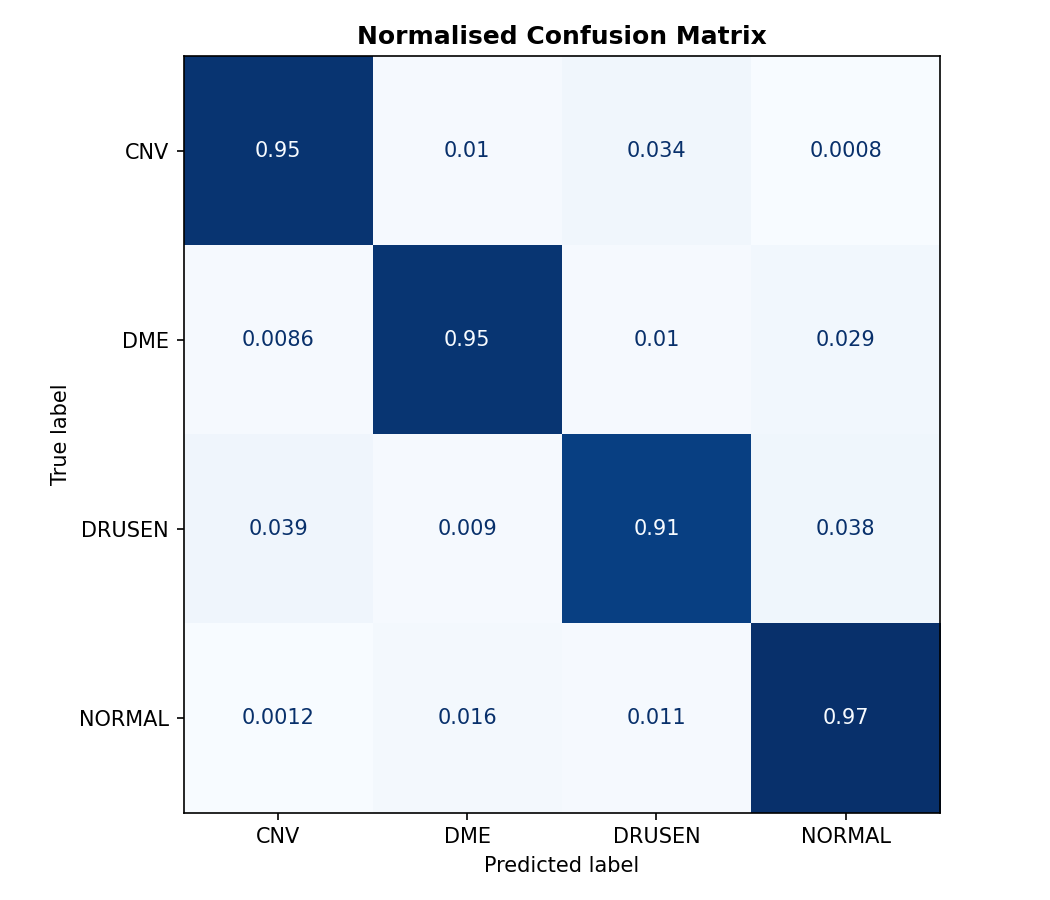}
\caption{Normalised confusion matrix on the test set (seed 42). Most
off-diagonal mass falls on the drusen row.}
\label{fig:cm}
\end{figure}

Figure~\ref{fig:gradcam} shows Grad-CAM and Transformer-attention overlays for
one scan per class: CNV activations concentrate on the choroidal membrane
beneath the RPE; DME activations mark intraretinal cyst pockets; drusen
activations fall on hyperreflective foci above the RPE; Normal scans produce
diffuse, low-magnitude maps with no focal pathology highlighted.
Figure~\ref{fig:cm} shows the normalised confusion matrix; most off-diagonal mass
falls on the drusen row, consistent with its lower training frequency.

\section{Discussion}
Accuracy across five seeds spans 95.16\%--95.70\%---narrow enough to be useful,
since a model that only reaches 95\% on lucky initialisations is not reliable
enough for a screening pipeline. ECE~$=0.0024$ means the model's stated
confidence tracks its actual accuracy, which matters more than raw accuracy when
a clinician is deciding whether to trust a result.

The main weakness is drusen F1~$=0.84$. Drusen lesions are diffuse and
low-contrast, and the $4.3\times$ imbalance versus CNV means the model sees
proportionally few drusen--normal boundary cases during training. Adding a
segmentation auxiliary head through multi-task learning is the most direct path
to improvement and is left for future work. A second limitation is
single-scanner scope: the Kermany dataset comes from one OCT device type, so
cross-vendor generalisation---which requires multi-site benchmarks such as
RETOUCH~\cite{bogunovic2019}---remains unvalidated here.

We also ran a parallel experiment fine-tuning RETFound~\cite{zhou2023}, a
ViT-L/16 foundation model pretrained on 1.6~million retinal images, on the same
dataset under an identical evaluation protocol. Full fine-tuning over 50 epochs
(three seeds) gave $95.14\%\pm1.37\%$ accuracy---comparable to our method but
with $5\times$ higher seed-to-seed variance. Our XGBoost head outperformed the
RETFound linear classification head specifically on drusen (F1 0.84 vs.\ 0.82),
suggesting the gradient-boosting decision surface remains valuable for the
minority class even when the encoder has been pretrained on domain data. A
frozen linear probe of RETFound reached only 75.72\% accuracy, confirming that
the modality gap between fundus photographs (RETFound's pretraining data) and OCT
B-scans is too large for a frozen ViT encoder to compensate.

\section{Conclusion}
We described a hybrid CNN--Transformer framework that reaches
$95.43\%\pm0.27\%$ accuracy on four-class retinal OCT classification alongside a
three-layer clinical safety envelope: temperature scaling (ECE 0.0024),
Mahalanobis OOD detection (97th-percentile threshold), and MC-Dropout
uncertainty quantification. The 2.07~GB Keras model compresses to a 237~MB ONNX
graph at $\approx63$~ms per scan on CPU, making it feasible for resource-limited
settings. To our knowledge, no prior OCT framework validates all three safety
mechanisms jointly with public weights and reproducible multi-seed evaluation.

\section*{Acknowledgements}
This work was conducted as an independent research project during the MSc
Advanced Computer Science programme at Newcastle University (2025--26). Compute
was provided through Google Colab. No external funding was received. The Kermany
OCT dataset was obtained from \url{https://data.mendeley.com/datasets/rscbjbr9sj/3}.

\section*{Data and Code Availability}
Source code, training notebook, and RETFound comparison experiments:
\url{https://github.com/Animesh-Kr/Human-Eye-Disease-Prediction}. Model weights,
Streamlit dashboard, Gradio pipeline, and REST inference API:
\url{https://huggingface.co/animeshakr/oct-retinal-weights}. The software is
archived on Zenodo at \href{https://doi.org/10.5281/zenodo.19224303}{DOI:
10.5281/zenodo.19224303}, linked via ORCID 0009-0003-0608-7004.

\section*{Competing Interests}
The author has no competing interests to declare. This paper describes a
decision-support system and does not replace professional ophthalmological
assessment.


\end{document}